\documentclass[%
 reprint,
superscriptaddress,
nofootinbib,
 amsmath,amssymb,
 amsfonts,
 aps,
]{revtex4-2}
\usepackage{graphicx}
\usepackage{enumitem}
\usepackage{macros}
\usepackage{dcolumn}
\usepackage{bm}
\usepackage{hyperref}
\hypersetup{
    colorlinks=true,
    linkcolor=blue,
    filecolor=blue,      
    urlcolor=blue,
    citecolor=blue, 
    pdftitle={Gibbs paper},
    pdfpagemode=FullScreen,
    }
\usepackage{appendix}
\usepackage[mathlines]{lineno}
\usepackage[mathlines]{lineno}
\usepackage[normalem]{ulem}
\usepackage{tabularx}
\newcolumntype{Y}{>{\centering\arraybackslash}X}
\usepackage[usenames,dvipsnames]{color}

\begin{document}

\preprint{APS/123-QED}

\title{Exploring the Capabilities of Gibbs Sampling in Pulsar Timing Arrays}
\author{Nima Laal}
\affiliation{Department of Physics, Oregon State University, 1500 SW Jefferson Way, Corvallis, OR 97331, USA}
 \email{nima.laal@gmail.com}
 
 \author{William G Lamb}
\affiliation{Department of Physics and Astronomy, Vanderbilt University, 2301 Vanderbilt Place, Nashville, TN 37235, USA}

\author{Joseph D. Romano}
\affiliation{Department of Physics, Texas Tech University, Box 41051, Lubbock, TX 79409, USA}

\author{Xavier Siemens}
\affiliation{Department of Physics, Oregon State University, 1500 SW Jefferson Way, Corvallis, OR 97331, USA}

\author{Stephen R. Taylor}
\affiliation{Department of Physics and Astronomy, Vanderbilt University, 2301 Vanderbilt Place, Nashville, TN 37235, USA}

\author{Rutger van Haasteren}
\affiliation{Max-Planck-Institut f{\"u}r Gravitationsphysik (Albert-Einstein-Institut), Callinstrasse 38, D-30167, Hannover, Germany}

\date{\today}
\begin{abstract}
We explore the use of Gibbs sampling in estimating the noise properties of individual pulsars 
and illustrate its effectiveness using the NANOGrav 11-year data set. We find that Gibbs sampling noise modeling (GM) is more efficient than the current standard Bayesian techniques (SM) for single pulsar analyses by yielding model parameter posteriors with average effective-sample-size ratio (GM/SM) of 6 across all parameters and pulsars. Furthermore, the output of GM contains posteriors for the Fourier coefficients that can be used to characterize the underlying red noise process of any pulsar's timing residuals, which are absent in current implementations of SM. Through simulations, we demonstrate the potential for such coefficients to measure the spatial cross-correlations between pulsar pairs produced by a gravitational wave background.
\end{abstract}

\maketitle


\section{\label{sec:Introduction}Introduction}
Pulsar timing arrays (PTAs)~\citep{Shazin,Detw} are low-frequency gravitational-wave (GW) detectors that use high-precision measurements of the times-of-arrival (TOAs) of pulses produced by an array of millisecond pulsars (MSPs). MSPs have ultra-stable spin periods on the order of milliseconds, and if their TOAs are measured to sufficient accuracy using large and sensitive radio telescopes, they can be used as cosmic clocks spread throughout our galaxy. Accurate models are constructed to predict the time at which each pulse is expected to arrive, and small deviations from the expected TOAs caused by GWs can be detected by searching for quadrupolar spatial correlations in those deviations between pulsars in the PTA~\cite{hd83}.

In recent years, multiple PTA searches for an isotropic stochastic gravitational wave background (GWB) have uncovered a common red noise process~\citep{12year,DR2, EPTA0, PPTA0}. 
This process was recently observed to posses a quadrupolar correlation signature matching the predictions of Einstein's general theory of relativity with various levels of significance~\citep{d1,d2,d3, d4}.

The sensitivity of PTAs to a GWB depends primarily on the number of pulsars in the array~\citep{ScalingLaws}. This is due to the fact that, at late times, the lowest frequencies in PTA data sets become GW-dominated, and the significance of the cross-correlations grows with the square root of the time span of the data and linearly with the number of pulsars in the array. In this regime, increasing the number of pulsars is the best way to maximize PTA sensitivity to the GWB. Currently, the International Pulsar Timing Array (IPTA) monitors 65 millisecond pulsars with 27 of such pulsars observed for more than 10 years~\citep{DR2}. For this reason, in each new release of a PTA data set the number of pulsars used in GWB detection analyses is expected to grow, which in turn makes the computational cost of noise modeling and parameter estimation increase significantly. This poses a significant challenge for Bayesian inference as typical searches for a GWB involve working with a very large parameter space making the use of computationally efficient algorithms a necessity. 

The standard Bayesian techniques for single and multi-pulsar noise modeling often result in a joint probability distribution for all of the model parameters (see \S\ref{sec:SM}). Despite the flexibility that this approach offers in choosing and implementing various noise models, the computational cost of parameter estimation using Markov Chain Monte Carlo (MCMC) simulations becomes prohibitive quickly. For instance, in the case of single-pulsar analyses, the number of parameters required to describe a pulsar's noise may well exceed forty (see \S\ref{sec:11Year}). This problem is more severe for the case of multi-pulsar analyses as even the simplest noise models require a number of parameters that is larger than twice the number of pulsars in the PTA. Hence, more computationally efficient data analysis techniques are critical for the future of PTA analyses.    

To mitigate these problems, there have been numerous efforts towards the development of more efficient Bayesian GWB detection techniques to analyze PTA data sets, such as those presented in~\citep{Lentati:2012xb,Gibbs0, HMC, rewe, BWM, parallel, fitting}. In particular, the work of \citet{Gibbs0} provides an outline for single-pulsar noise analyses in which Gibbs sampling can be used to characterize the red noise component of each pulsar's timing residuals. In this paper, we explore the capabilities of the Gibbs sampling method in single-pulsar noise analyses by applying it on the NANOGrav 11-year data set~\citep{11yr} as well as simulated data sets. We show that the Gibbs sampling method is well suited for PTA single-pulsar analyses and results in probability distribution functions for all model parameters in a significantly shorter time-scale compared to those obtained via the standard MCMC methods. Furthermore, we show, via simulated data sets, that the Fourier coefficients that result from the Gibbs sampling procedure can be used to identify the shape of the underlying spatially-correlated signal in a PTA data set. 

The paper is structured as follows. In \S\ref{sec:Methods}, we review and simplify the methods presented in \citet{Gibbs0} to outline the Gibbs sampling method and its accompanying noise modeling. Furthermore, in order to use the output of Gibbs sampling in a subsequent multi-pulsars analysis, and inspired by~\citet{OPTSTAT0}, we introduce our version of a frequency domain optimal statistic which follows from the PTA multi-pulsar likelihood function. In \S\ref{sec:11Year}, we employ the outlined method in order to analyze the NANOGrav 11 year data set and compare the results to those obtained by the standard Bayesian PTA detection techniques. Finally, in \S\ref{sec:sims}, we analyze PTA simulated data sets to reveal the potential of the Gibbs sampling technique in searches for a common correlated signal across an array of pulsars. 
\begin{table*}
\begin{tabular}{cc}
\hline
\textbf{Symbol} & \textbf{Description}                                               \\ \hline
$T_{\text{obs}}$               & Observational baseline                                            \\
$t$               & Time
\\
$f$               & Frequency                                                         \\
$I$, $J$            & Pulsar indices                                                   \\
$k$, $s$               & Indices for the frequency bins                                          \\
$m$               & Index for the number of pulsars in the array
                \\
$p$               & Number of TOAs for a given pulsar
                \\
$q$               & Number of timing model parameters
                \\
$\bm{r}$               & Timing residual                                                   \\ \hline
$F$               & Fourier design matrix                                            \\
$M$               & Timing model design matrix                                                     \\
$T$               & Combination of $F$ and $M$ such that $T=[M, F]$ 
\\
$N$               & White noise covariance matrix                                     \\
$B$               & Covariance matrix for the linear timing model parameters and the Fourier coefficients (i.e., $\langle \bm{b} \bm{b}^T \rangle$)           \\
$\Gamma$               & Hellings and Downs cross correlation matrix           \\
$\bm{A}$               & Collection of Fourier coefficients across pulsars and frequencies $ \{ a_{k;m}\}$,            \\
$\varphi$          & Single-pulsar red process covariance matrix                                      \\ 
$\Phi$          & Multi-pulsar red process covariance matrix
                \\
\hline
$\bm{a}$               & Fourier sin-cos coefficients 
\\
$\lambda$               & Estimated cross correlations           \\
$ \bm{\epsilon}$          & Linear timing model parameters
\\
$\bm{b}$               & Combination of $\bm{a}$ and $ \bm{\epsilon}$ such that $\bm{b}^T=[ \bm{\epsilon}, \bm{a}]$ 
\\
$\bm{w}$               & White noise time series              
\\
$\bm{n}$               & Collection of all white noise parameters             
\\
$\bm{\rho}$           & Free-spectrum parameter used in describing power-spectral-density ($\bm{\rho}^2 =\langle \bm{a} \bm{a}^T \rangle$) \\
$A$           & Amplitude of a red noise process \\
$P$           & Power-spectral-density of a red noise process \\
$\hat{P}$           & Spectral shape of a red noise process obtained by $P/A^2$ \\ \hline
\end{tabular}
\caption{A table listing the symbols most commonly used throughout this paper and a short description of what they represent. Refer to \S\ref{sec:terminology} for more details on the definition of some of the quantities.}
\label{params}
\end{table*}
\section{\label{sec:Methods}Methods}
We begin our review of the Gibbs sampling method \citep{Gibbs0} by writing a simple model for a pulsar's post-fit timing residuals, $\bm{r}$, in terms of a set of Fourier coefficients $\bm{a}$, Fourier design matrix $F$, linear timing model parameters $ \bm{\epsilon}$, timing design matrix $M$, and white noise $\bm{w} $\footnote{Refer to \autoref{params} and \S\ref{sec:terminology} for more details on the definitions of the quantities used throughout this paper.}:
\begin{align} \label{firsteq}
\begin{split}
{\bm{r}}&=M \bm{\epsilon} +F\bm{a}+\bm{w} \\ &=T \bm{b} + \bm{w},
\end{split}
\end{align}
where $\bm{b}^T=[ \bm{\epsilon}, \bm{a}]$ and $T = [M, F]$. Assuming Gaussian white noise, parameterized by the set of parameters $\bm{n}$ with prior $p(\bm{n})$, the above model allows for the construction of posterior probability density functions following Bayes' theorem:
\begin{align}
p\left( \left. \bm{\rho} ,\bm{b}, \bm{n} \right|\bm{r} \right) \propto p\left( \left. \bm{r} \right|\bm{b}, \bm{n} \right)p\left( \bm{a} \right) p\left( \bm{\epsilon} \right) p\left( \bm{\rho}  \right)p\left( \bm{n} \right),
\end{align}
where,
\begin{align}
p\left( \left. \bm{r} \right|\bm{b}, \bm{n} \right)&=\frac{\exp \left\{ -\frac{1}{2}\left[ \left( \bm{r}-T\bm{b} \right)^{T}{{N}^{-1}}{{\left( \bm{r}-T\bm{b} \right)}} \right] \right\}}{\sqrt{\det \left\{(2\pi) N \right\}}} \label{liklihood}, \\ 
p\left( \bm{a} \right)&=\frac{\exp \left\{ -\frac{1}{2}\left[ \bm{a}^{T}\varphi^{-1} {{\bm{a}}} \right] \right\}}{\sqrt{ \det \left\{(2\pi) \varphi \right\}}} \label{a-prior}, \\ 
 p\left( \bm{\rho}  \right)&=\prod\limits_{s=1}^{k}{\frac{1}{\rho_s}}, \label{rho-prior}
\end{align}
for 
\begin{align}
 \varphi &=\left\langle {\bm{a} \bm{a}^T} \right\rangle, \\
 B&=\left\langle {{\bm{b}}}\bm{b}^{T} \right\rangle ,
\end{align}
and $\bm{\rho}$ denoting the collective set $\left\{ {{\rho }_{1}},{{\rho }_{2}},\ldots ,{{\rho }_{k}} \right\}$ whose elements are used to parameterize a pulsar's power-spectral-density, frequency-bin by frequency-bin\footnote{Note that the total number of frequency-bins is $k$, but there are two Fourier coefficients per each frequency-bin. Both $a_s^\text{cos}$ and $a_s^\text{sin}$ have the same variance parameterized by $\rho_s^2$. This is reflected in \autoref{a-prior}.}, and describe the variance of the Fourier coefficients. Additionally, a log-uniform (conjugate) prior $p(\rho_k)=1/\rho_k$ is considered as seen in \autoref{rho-prior}. 

Moreover, we have assumed an unbounded improper prior for the linear timing model parameters and have set $\langle \bm{\epsilon} \bm{\epsilon}^T \rangle  = \text{diag} \{\infty\}$. Such choices for the linear timing model parameters are typical of PTA noise analyses due to the lack of physically-motivated priors for all of the timing model parameters and are acceptable as long as the data is informative with respect to such parameters. Hence, we can write
\begin{align}
       B^{-1}&=\left[ \begin{matrix}
   0 & 0  \\
   0 & {{\varphi }^{-1}}  \\
\end{matrix} \right] \label{BINV}.
\end{align}

To proceed with Gibbs sampling, the posterior for each of the model parameters needs to be cast into a conditional probability distribution form where each model parameter is conditioned upon the other model parameters and the timing residuals. In the following two subsections, we derive such conditional probabilities for parameters $\bm{b}$ and $\bm{\rho}$. 
\subsection{\label{a}Conditional probability of coefficients}
For the coefficients $\bm{b}$, the conditional probability can be found by rewriting the full posterior (i.e., the product of \autoref{liklihood}, \autoref{a-prior} and \autoref{rho-prior}) while ignoring all factors not depending on $\bm{b}$ coefficients explicitly. In other words, all model parameters are treated as constants and only the $\bm{b}$ coefficients are allowed to vary:
\begin{align}
\ln p\left( \left. \bm{b} \right|\bm{\rho} ,\bm{r},\bm{n} \right) \simeq &-\frac{1}{2}\big[ \left( \bm{r}-T\bm{b} \right)^{T}{{N}^{-1}}{{\left( \bm{r}-T\bm{b} \right)}} \nonumber \\ &+ \bm{b}^{T} B^{-1} {\bm{b}} \big] \nonumber \\
\begin{split}\label{logb}
=&-\frac{1}{2}\left[ {\bm{b}^{T}}\left( {{T}^{T}}{{N}^{-1}}T+{{B }^{-1}} \right)\bm{b}\right] \\ &-\frac{1}{2} \left[-2{\bm{b}^{T}}{{T}^{T}}{{N}^{-1}}\bm{r} \right].
\end{split}
\end{align}
The above equation suggests that the $\bm{b}\text{-dependence}$ of the probability $p\left( \left. \bm{b} \right|\bm{\rho} ,\bm{r},\bm{n} \right)$ is Gaussian. Using the \emph{maximum a posteriori} estimate of $\bm{b}$ found by maximizing \autoref{logb} as an estimate of the mean of the Gaussian, one can write the conditional probability distribution of the $\bm{b}$ coefficients, \autoref{logb}, in the form
\begin{align}
p\left( \left. \bm{b} \right|\bm{\rho} ,\bm{r},\bm{n} \right)=\frac{\exp \left\{ -\frac{1}{2}{{\left( \bm{\hat{\mu}} -\bm{b} \right)}^{T}}{{\Sigma }}\left( \bm{\hat{\mu}} -\bm{b} \right) \right\}}{\sqrt{ \det \left\{(2\pi) \Sigma^{-1}  \right\}}}, \label{bgivenrho}
\end{align}
where \footnote{The definition of $\Sigma$ in \autoref{sigma_Def} is chosen so that this paper's $\Sigma$ represents the same quantity as the $\Sigma$ defined in the PTA GWB detection literature.},
\begin{align}
 \Sigma&={{T}^{T}}{{N}^{-1}}T+{{B}^{-1}} \label{sigma_Def}, \\ 
 \hat{\mu}&={{\Sigma^{-1} }}{{T}^{T}}{{N}^{-1}}\bm{r} \label{mudef}.
\end{align}
\subsection{\label{rho}Conditional probability of red noise power-spectral-density}
Similar to the $\bm{b}$ coefficients, the conditional probability of the $\bm{\rho}$ parameters can be found by taking advantage of the full posterior and ignoring all the factors not depending on $\bm{\rho}$ explicitly. Additionally, we make the observation such that the relevant probability distributions can be factorized over frequency-bins:
\begin{align}
   p\left( \left. \bm{\rho}  \right|\bm{a},\bm{r},\bm{n} \right)&=\prod\limits_{s=1}^{{k}}{p\left( \left. {{\rho }_{s}} \right|{{a}_{s}},\bm{r},\bm{n} \right)} \nonumber \\ 
 & =\prod\limits_{s=1}^{{k}}{\frac{1}{{{\rho }_{s}}\sqrt{(2\pi) \rho _{s}^{2}}}\exp \left\{ -\frac{1}{2}\left( \frac{{{a}_{s}}\cdot {{a}_{s}}}{{{\rho }_{s}}} \right) \right\}} \nonumber \\ 
 & \propto \prod\limits_{s=1}^{{k}}{\frac{1}{\rho _{s}^{2}}\exp \left\{ -\left( \frac{\frac{{{a}_{s}}\cdot {{a}_{s}}}{2}}{{{\rho }_{s}}} \right) \right\}}\nonumber \\ 
 & =\prod\limits_{s=1}^{{k}}{\text{InvGamma}\left( \alpha =1,\beta =\frac{{{a}_{s}}\cdot {{a}_{s}}}{2} \right)}. \label{rhogivenb}   
\end{align}
In the above, the dot-product denotes the sum of the square of the cosine and sine Fourier coefficients for each frequency-bin that is ${a}_{s}\cdot {a}_{s} = (a_{s}^\text{cos})^2 + (a_{s}^\text{sin})^2$ . Furthermore, despite the analytic form for the dependence of $\bm{\rho}$ on the Fourier coefficients $\bm{a}$, the lower and the upper bounds of the inverse-gamma distribution extending to zero and infinity would lead to astrophysically and statistically incorrect assumptions as such bounds need to be finite and constrained to avoid the implicit use of improper priors in the modeling of red noise processes. Thus, a truncated version of the derived inverse-gamma distribution needs to be considered. In \S\ref{truncinvgamma}, we show how to obtain such a truncated distribution. 
\subsection{\label{whitenoise}Conditional probability of white noise parameters}
In contrast to $\bm{b}$ and $\bm{\rho}$, the white noise parameters cannot be written in terms of standard statistical distributions. This is mainly due to the dependence of the white noise parameters to various radio telescope receivers (i.e., each backend of each radio telescope needs its own white noise parameters). Solving the full-likelihood for the white noise parameters, collectively denoted by $\bm{n}$, results in 
\begin{align}
\begin{split}\label{white-prob}
     \ln p\left( \left. \bm{n} \right|\bm{\rho} ,\bm{b},\bm{r} \right)=&-\frac{1}{2}\sum\limits_{i=1}^{p}{\left\{ {{\left( \bm{r}-T\bm{b} \right)}^{T}}{{N}^{-1}}\left( \bm{r}-T\bm{b} \right) \right\}} \\
  &-\frac{1}{2}\sum\limits_{i=1}^{p}{\ln \left( \det{ \left \{2\pi N \right \}} \right)},
  \end{split}
\end{align}
where the sum is over the TOAs. Since \autoref{white-prob} cannot be simplified further in any useful way, we have no choice but to utilize a non-Gibbs MCMC procedure to sample the posterior. 
\subsection{\label{sec:SM}Standard method of single-pulsar analyses}
The standard method of single-pulsar analyses involves an analytical marginalization of the product of \autoref{liklihood} and \autoref{a-prior} over the the coefficients $\bm{b}$. The result is
\begin{align}
     p\left( \left. \bm{r} \right|\bm{\rho}  \right)&=\frac{1}{\sqrt{ \text{det} \left\{{{\left( 2\pi  \right)}}C \right\}}}\exp \left\{ -\frac{1}{2}{\bm{r}^{T}}C^{-1}\bm{r} \right\} \label{SMEQ}, \\ 
  C&=N+{{T}}BT^{T}, \\ 
{{C}^{-1}}&={{N}^{-1}}-{{N}^{-1}}T{{\Sigma}^{-1}}{{T}^{T}}N^{-1}, 
\end{align}
where in the last line, we have used the \emph{Woodbury identity}:
\begin{align}
    &{{\left( X+UYV \right)}^{-1}}= \nonumber \\
    &{{X}^{-1}}-{{X}^{-1}}U{{\left( {{Y}^{-1}}+V{{X}^{-1}}U \right)}^{-1}}V{{X}^{-1}},
\end{align}
and $\Sigma$ is defined in \autoref{sigma_Def}.
The dependence of \autoref{SMEQ} on the red noise parameters $\rho_k$ is through the elements of the matrix $\Sigma^{-1}$. Once \autoref{SMEQ} is multiplied by the appropriate priors of the model parameters, the resulting joint probability distribution of $ p\left( \left. \bm{\rho} \right|\bm{r}  \right)$ is ready to be given to a non-Gibbs MCMC algorithm for parameter estimation.
\subsection{\label{Gibbs}Gibbs sampling}
\begin{figure}
\includegraphics[width=\linewidth]{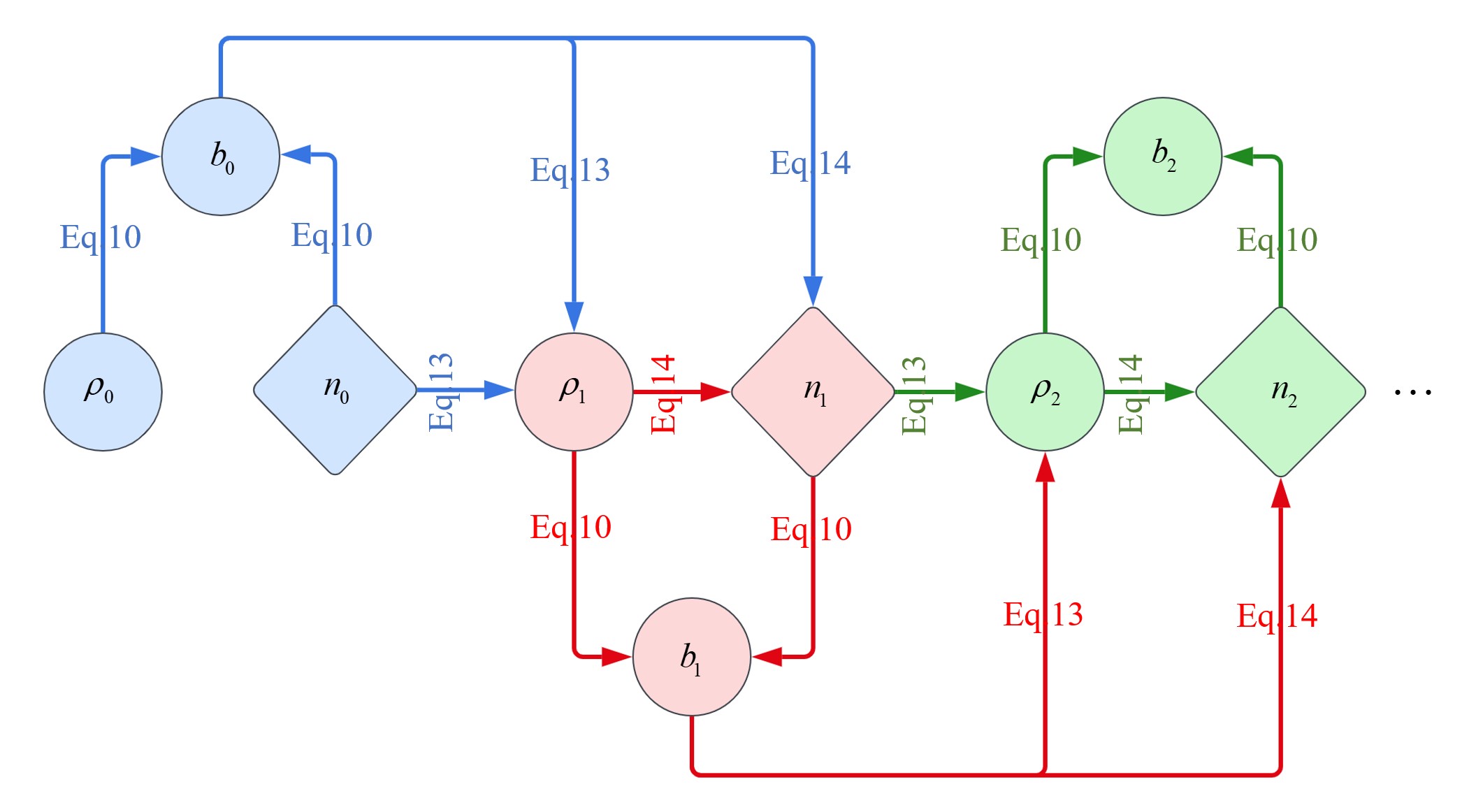}
    \caption{A schematic representation of the first three steps of the outlined Gibbs sampling procedure. The first step of the sampling process (blue) starts by guesses of the $\bm{\rho}$ and the white noise parameters and results in an estimate of the coefficients $\bm{b}$ following \autoref{bgivenrho} using the previously guessed values. The second (red) and the third (green) steps of the sampling continue the sequence by estimating the next remaining model parameter given the most recent estimates of the other two parameters using the conditional probability distributions of \autoref{bgivenrho}, \autoref{rhogivenb}, and \autoref{white-prob}.}
    \label{Gibbs Sampling}
\end{figure}

Gibbs sampling \citep{Metropolis1953} is a MCMC algorithm designed to take advantage of the conditional probability distributions of all model parameters in order to perform parameter estimation. It is often used in statistical inferences where a joint probability distribution of all parameters is difficult to sample, yet each model parameter's probability distribution can be written in terms of the rest of the parameters and the data. Gibbs sampling allows for random draws from the conditional probability distributions of model parameters whose analytic functional form must be found prior to the start of the sampling process as we have done for the case of single-pulsar noise analyses by deriving \autoref{bgivenrho} and \autoref{rhogivenb}. Due to the existence of analytic forms for the probabilities, the concept of rejection of random states, an integral part of the other MCMC algorithms, does not belong to the Gibbs sampling as all draws are considered accepted. Nevertheless, Gibbs sampling is still a MCMC algorithm as it possesses features such as no long-term-memory and the need for burn-in of the final Markovian chain. We will outline a step-by-step implementation of Gibbs sampling for a single-pulsar noise analysis in the remaining part of this section. 

Knowing the conditional probabilities of our model parameters, $\bm{\rho}$, $\bm{b}$, and $\bm{n}$, it is simple to implement Gibbs sampling in the following way:
\begin{itemize}[leftmargin=*]
\item[]\textbf{Step 1}: Make initial guesses of $\bm{\rho}$ and $\bm{n}$ denoted by $\bm{\rho}_0$ and $\bm{n}_0$.
\item[]\textbf{Step 2}: Using \autoref{bgivenrho}, find an estimate of $\bm{b}_0$ given $\bm{\rho}_0$ and $\bm{n}_0$.
\item[]\textbf{Step 3}: To start the first iteration, find an estimate of $\bm{\rho}_1$ given $\bm{b}_0$ and $\bm{n}_0$ using \autoref{rhogivenb}.
\item[]\textbf{Step 4}: Continuing the first iteration, find an estimate of $\bm{n}_1$ given $\bm{b}_0$ and $\bm{\rho}_1$ with a very short MCMC procedure sampling \autoref{white-prob}.
\item[]\textbf{Step 5}: To end the first iteration, find an estimate of $\bm{b}_1$ given $\bm{\rho}_1$ and $\bm{n}_1$ using \autoref{bgivenrho}.
\end{itemize}
 
 \autoref{Gibbs Sampling} provides an illustration of the explained procedure. The above steps can be repeated until all the model parameters reach satisfactory convergence. Due to the analytical draws of the $\bm{\rho}$ and the $\bm{b}$ coefficients, convergence will be reached quickly compared to the fully non-Gibbs MCMC algorithms. This is one of the most desirable features of Gibbs sampling as the overall run-time of the PTA single-pulsar noise analyses will be reduced significantly.

 
\subsection{\label{sec:M3A} Frequency domain multi-pulsar likelihood}
The outlined Gibbs sampling procedure is an efficient Bayesian scheme capable of estimating each pulsar's power-spectral-density as well as the Fourier coefficients required to describe the total red noise (i.e., GWB plus spatially-uncorrelated intrinsic red noise process) component of the timing residuals. However, the information required in characterising a GWB requires subsequent multi-pulsar analyses. As will be demonstrated in this section, the output of Gibbs sampling provides enough information to perform multi-pulsar analyses aiming at detecting a GWB. 

Using only the Fourier coefficients $\bm{a}$, one can construct a factorized likelihood in the frequency domain in the following way:
\begin{align}
   p\left( \left. \bm{A} \right|  \Phi \right)&={\frac{\exp \left\{ -\frac{1}{2}\left( a_{k;I}^T\Phi _{ks}^{-1}{{a}_{s;J}} \right) \right\} }{\sqrt{ \det \left\{{{(2\pi)\Phi }} \right\}}}}  \label{M3A}\\
   &= \prod\limits_{s} {\frac{\exp \left\{ -\frac{1}{2}\left( a_{s;I}^T\Phi _{ss}^{-1}{{a}_{s;J}} \right) \right\} }{\sqrt{ \det  \left\{{{(2\pi)\Phi_{ss} }} \right\}}}},
\end{align}
for 
\begin{align}
    \Phi &={{\left[ \begin{matrix}
   {{{{\Phi }}}_{11}} & 0 & \ldots  & 0  \\
   0 & {{{{\Phi }}}_{22}} & \ldots  & 0  \\
   \vdots  & \vdots  & \ddots  & \vdots   \\
   0 & 0 & \ldots  & {{{{\Phi }}}_{kk}}  \\
\end{matrix} \right]}}, \\
   {{\Phi }_{kk}}&={{\left[ \begin{matrix}
   \varphi_{k,1} & {{\Gamma }_{k;12}}\rho _{k;g}^{2} & \ldots  & {{\Gamma }_{k;1m}}\rho _{k;g}^{2}  \\
   {{\Gamma }_{k;21}}\rho _{k;g}^{2} & \varphi_{k,2} & \ldots  & {{\Gamma }_{k;2m}}\rho _{k;g}^{2}  \\
   \vdots  & \vdots  & \ddots  & \vdots   \\
   {{\Gamma }_{k;m1}}\rho _{k;g}^{2} & {{\Gamma }_{k;m2}}\rho _{k;g}^{2} & \ldots  & \varphi_{k,m}  \\
\end{matrix} \right]}}, \nonumber\\
\end{align}
where $\bm{A}$ denotes the collection of Fourier coefficients, across all pulsars and all frequencies, (i.e., $\bm{A} = \{ a_{k;m}\}$), $\rho_{k;g}$ parameterizes the common power-spectral-density of the GWB (indicated by the subscript $g$) observed across the entire pulsar array at frequency $k$, and $\Gamma_{IJ}$ represents the functional form of the cross correlations (e.g., Hellings and Downs curve).

One can use this equation to derive an optimal estimator of the signal-to-noise analogous to those presented in \citet{OPTSTAT0} and \citet{TimeDomainPaper}. We leave the details of the derivation to our future project \citep{DD} where we explore the use of the Fourier coefficients in GWB characterisation in great detail. Here, we simply report the results in the form of the optimal estimators of the cross-correlations $\lambda_{IJ}$ and their uncertainty $\sigma_{IJ}$:
\begin{align}
{{\lambda }_{IJ}}&=\frac{\sum\limits_{s}{{{a}_{s;I}} \cdot {{a}_{s;J}}{{}}\frac{{\hat{P}_g}}{{\varphi_{s;I}}{\varphi_{s;J}}}}}{\sum\limits_{s}{\frac{{{{\hat{P}_g^2}}}}{{\varphi_{s;I}}{\varphi_{s;J}}}}}\label{lambda}, \\ 
{{\sigma }_{IJ}}&={{\left[ \sum\limits_{s}{{ {}}\frac{{\hat{P}_g^2}}{{\varphi_{s;I}}{\varphi_{s;J}}}} \right]}^{-\frac{1}{2}}} \label{lambdasigma}. 
\end{align}

Without a need for a detailed derivation, \autoref{lambda} and \autoref{lambdasigma} can be understood by following a very simple rational. The numerator is the weighted product of $a_{s;I} \cdot a_{s;J}$. The weights associated with such product, $1/\varphi_I$ and $1/\varphi_J$, have the role of suppressing the contributions from pulsars whose total non-GWB noise power is substantial (i.e., dominant spatially-uncorrelated intrinsic red noise). Moreover, the choice for the normalization in the denominator ensures that the estimated correlations would yield $A_g^2 \Gamma_{IJ}$ if averaged over many realizations of GWB as is shown in \S\ref{FDOSNORM}.
\begin{figure*}[t]
    \includegraphics[width=\linewidth]{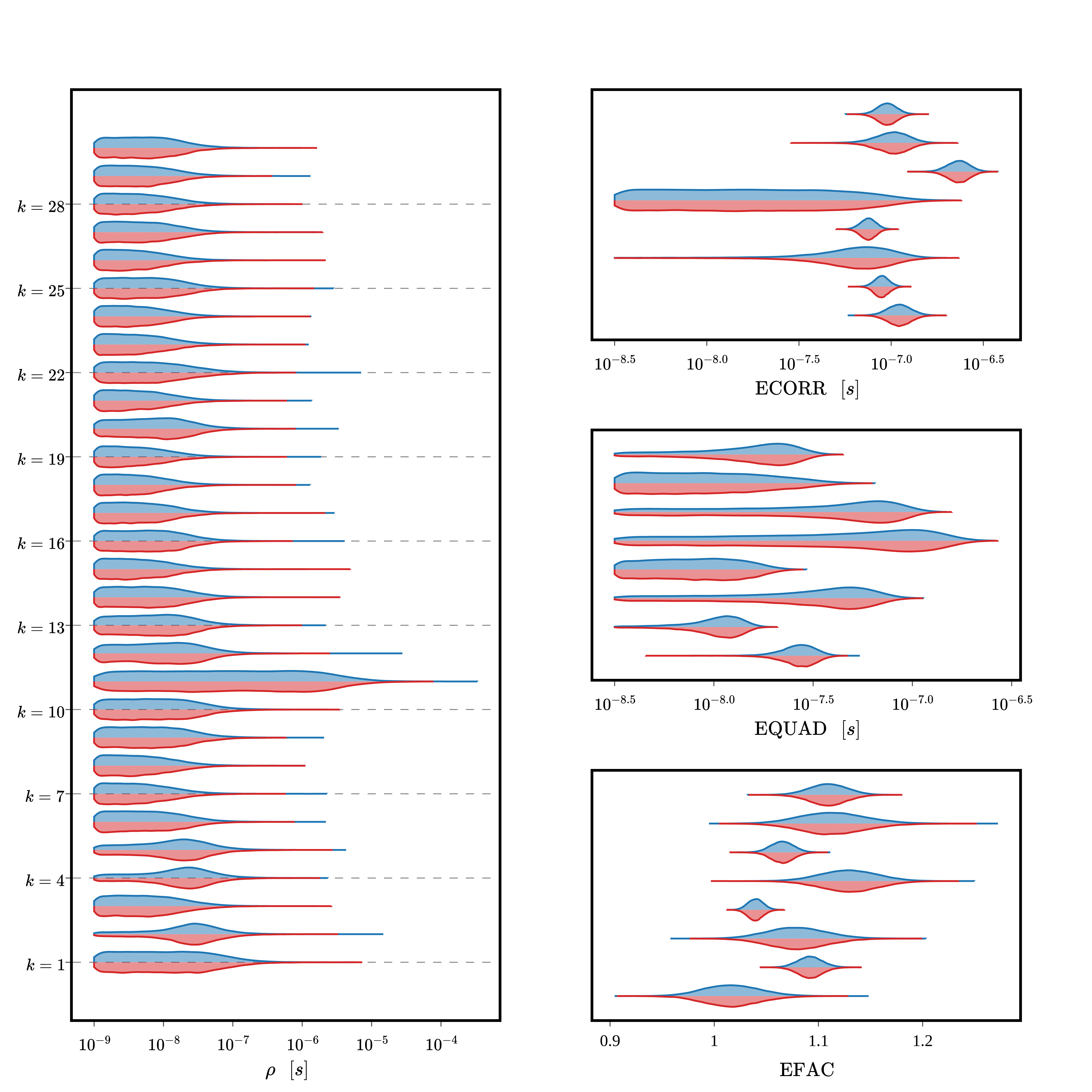}
    \caption{A comparison of posteriors for all model parameters of PSR J$1713+0747$ for the NANOGrav 11 year data set obtained via GM (blue) and SM (red). The posteriors on the left column belong to the red noise model parameters, collectively referred to as ${\bm{\rho}}$, whereas the posteriors on the right column belong to the white noise parameters EFAC, EQUAD, and ECORR. There is one ${{\rho_k}}$ parameter for each frequency ($k=30$ frequencies in total) and three white noise parameters for each receiver (8 receivers in total). To obtain the plots via GM, 30 steps of a Metropolis Hasting algorithm within each step of Gibbs sampling has been implemented for the white noise parameters. The above plots show a great level of consistency in extracting the posteriors between the two methods.}
    \label{POS Comp}
\end{figure*}

Additionally, estimates of the amplitude, the uncertainty of the estimated amplitude, and the signal-to-noise-ratio (SNR) can be made from \autoref{lambda} and \autoref{lambdasigma} by minimizing a weighted-chi-squared statistic of the form
\begin{align}
    {{\chi }^{2}}=\sum\limits_{IJ}{\frac{{{\left( {{\lambda }_{IJ}}-A_{g}^{2}\Gamma _{IJ} \right)}^{2}}}{\sigma _{IJ}^{2}}},
\end{align}
with respect to $A_g^2$ which results in 
\begin{align}
{{{\hat{A}}}_{g}^{2}}&=\frac{\sum\limits_{IJ;I\ne J}{\sum\limits_{s}{{{a}_{s;I}}\cdot{{a}_{s;J}}{{\Gamma }_{IJ}}\frac{\hat{P}_g}{{\varphi_{s;I}}{\varphi_{J;s}}}}}}{\sum\limits_{IJ;I\ne J}{\sum\limits_{s}{{{\Gamma }_{IJ}^2}\frac{\hat{P}_g^2}{{\varphi_{I;s}}{\varphi_{s;J}}}}}}, \label{opt} \\ 
{{\sigma }_{g}}&={{\left[ \sum\limits_{IJ;I\ne J}{\sum\limits_{s}{{{\Gamma }_{IJ}^2}\frac{\hat{P}_g^2}{{\varphi_{I;s}}{\varphi_{J;s}}}}} \right]}^{-\frac{1}{2}}}, \\
\text{SNR} &= \frac{\hat{A}_g^2}{\sigma_g} \label{SNR}.
\end{align}

When estimating the optimal correlations using \autoref{lambda}, one has a few options to select from for the choice of $\bm{a_I}$ and $\bm{a_J}$. The trivial option is to draw randomly from the multivariate probability distribution of each pulsar's $\bm{a}$ (the output of Gibbs sampling) and obtain the cross product of such random draws for each pulsar pair. Another option is to construct posteriors of the mean, $\hat{\mu}$, following \autoref{bgivenrho}, and draw randomly from such posteriors. Similar to the previous case, the cross product of the random draws can be used in \autoref{lambda} and with the difference that the normalization factor in the denominator of \autoref{lambda} should be recalculated (see \S\ref{FDOSNORM} for more details). Lastly, for the choice of $\varphi_{I}$, we use the total red noise power $P_I$.

As a final note, it is important to recognize the limitations of the presented technique as well as the \emph{optimal statistic} in general. In practice, optimal statistic results in biased estimates of the GWB amplitude and the signal-to-noise ratio if one does not have separate estimates for the spatially-uncorrelated as well as the common red noise power. In other words, if one uses the red noise power estimates from the single-pulsar analyses instead of obtaining separate estimates for a common red noise signal and intrinsic red noise signal, one cannot characterise a common correlated signal correctly. This has been explored in depth in \citep{NMOS}. 
\begin{figure*}[!tbp]
  \centering
  \begin{minipage}[t]{0.49\linewidth}
    \includegraphics[width=\linewidth]{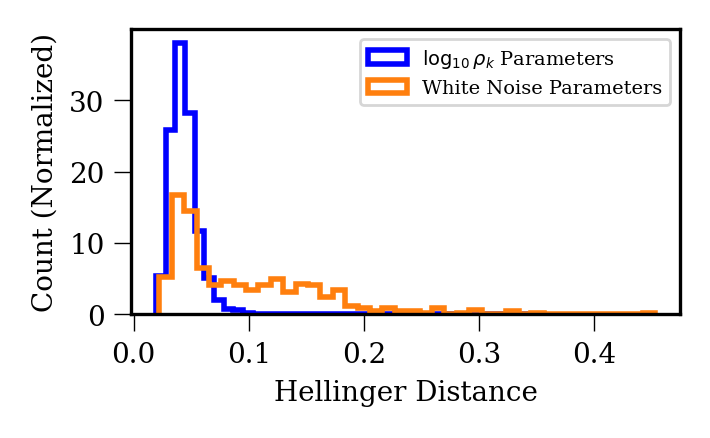}
    \caption{A histogram showcasing the distribution of the Hellinger distance values for the $\log_{10}{\bm{\rho}}$ posteriors (blue) and the white noise parameters (orange) obtained by comparing the outputs of GM and SM. The histogram contains the Hellinger distances of model parameters across all frequencies and pulsars. As evident by the distribution, GM and SM result in sufficiently similar distributions with a few exceptions whose inconsistencies can be attributed to the differences in the level of convergence of posteriors resulting from GM and SM even though we have allowed sufficient time for SM to converge (i.e., more than two hours). GM posteriors follow the general shape of SM posteriors but are significantly more converged.}
    \label{HELL1}
  \end{minipage}
  \hfill
  \begin{minipage}[t]{0.49\linewidth}
    \includegraphics[width=\linewidth]{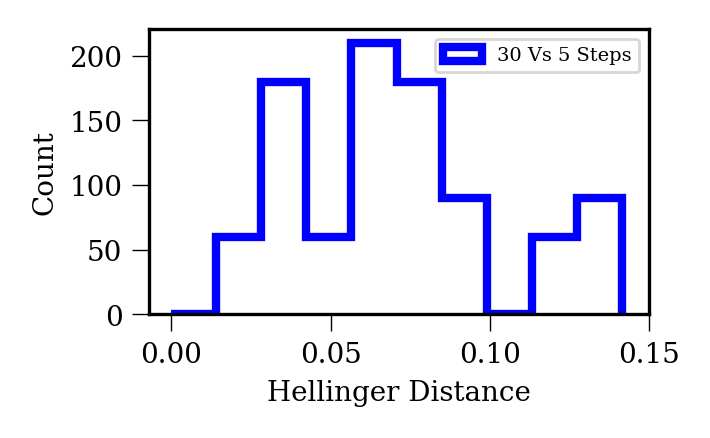}
    \caption{A histogram showcasing the distribution of the Hellinger distance values between the $\log_{10}{\bm{\rho}}$ posteriors obtained using GM with two different number of MCMC steps (30 and 5 steps) for each step of Gibbs sampling for the white noise parameters. The figure is made by combining the Hellinger distance values across all of the pulsars and all of the frequencies. As evident by the distribution, choosing a much lower number of MCMC steps for each step of Gibbs sampling for the white noise parameters does not change the shape of the target $\log_{10}{\bm{\rho}}$ posteriors significantly. This effect can be attributed to the knowledge of GM about the analytical shape of the $\log_{10}{\bm{\rho}}$ parameters prior to the start of the sampling.}
    \label{HELL2}
  \end{minipage}
\end{figure*}
\section{\label{sec:11Year}Analysis of the NANOGrav 11 Year Data Set}
To test the capabilities of the outlined single-pulsar data analysis technique, we analyze the NANOGrav 11 year data set \citep{11yr} using Gibbs sampling. The results are then compared to the ones obtained via standard Bayesian modeling detection routine used by the NANOGrav collaboration in their most recent work \citep{12year}. To ensure the fairness of the convergence comparisons, we allow each technique to sample the data set for two hours for each pulsar. After the two hours time-limit, we compare the posteriors' effective-sample-size (ESS) and rank-normalized-split R-hat ($\hat{r}$) values using the diagnostic tools provided by \citet{arviz_2019}.
\subsection{\label{sec:Modeling}Details of the Bayesian modeling}
The Gibbs sampling implementation used for the 11 year data set models the data as outlined in \S\ref{sec:Methods}. This Bayesian modeling together with Gibbs sampling is referred to as Gibbs Method (GM) from hereon. Moreover, the competing method of analyzing the NANOGrav 11 year data set follows the standard single-pulsar analyses currently implemented in the most recent GWB searches \citep{d1, d2, d3} and explained in \S\ref{sec:SM}. The PTMCMC sampling package \citep{PTMCMC} as well as the structure of the Bayesian modeling accompanying this sampling is referred to as Standard Method (SM) from hereon. 

For both SM and GM, we have allowed each pulsar's set of red noise parameters, $\bm{\rho}$, to follow a 30 frequency free-power-spectral-density model with frequencies ranging from $1/T_{\text{obs}}$ to $30/T_{\text{obs}}$ in which $T_{\text{obs}}$ denotes the observational baseline of each considered pulsar. The choice of prior for the model parameters are listed below. For each pulsar, the white noise parameters are per receiver/backend system while the $\bm{\rho}$ parameters are per frequency: 
\begin{align}
    \bm{\rho} \quad [s] &\sim \text{log-Uniform}(-9, -4) \label{lur}, \\
    \text{EQUAD} \quad [s] &\sim \text{log-Uniform}(-8.5, -5)\label{lueq}, \\
\text{ECORR} \quad [s] &\sim \text{log-Uniform}(-8.5, -5)\label{luec}, \\
    \text{EFAC} &\sim \text{Uniform}(0.01, 10)\label{uef},
\end{align}
for $[s]$ denoting the unit of the quantities, which is seconds.
\begin{figure*}[!tbp]
  \centering
  \begin{minipage}[t]{0.49\linewidth}
    \includegraphics[width=\linewidth]{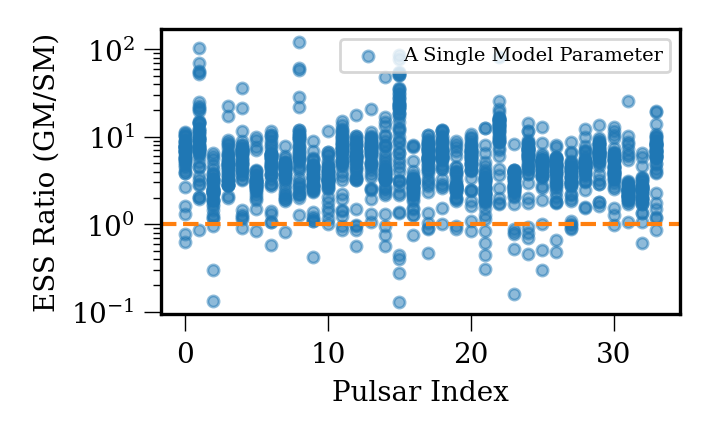}
    \caption{A scatter-plot showcasing the differences in the spread of the effective-sample-size (ESS) values for the $\log_{10}{\bm{\rho}}$ and the white noise parameters expressed in the form of the ratio of GMs' ESS over SMs' ESS (blue circles). For each pulsar, there is one $\log_{10}{\bm{\rho_k}}$ for each frequency (30 frequencies in total) and three white noise parameters for each receiver. Across all of the pulsars, GM is more capable at yielding posteriors with significantly higher ESS levels given the two hour time limit. Considering all model parameters, the average ESS ratio is 6. The values of ESS are found using the functionalities provided in \citet{arviz_2019}.}
    \label{ESS Comp}
  \end{minipage}
  \hfill
  \begin{minipage}[t]{0.49\linewidth}
    \includegraphics[width=\linewidth]{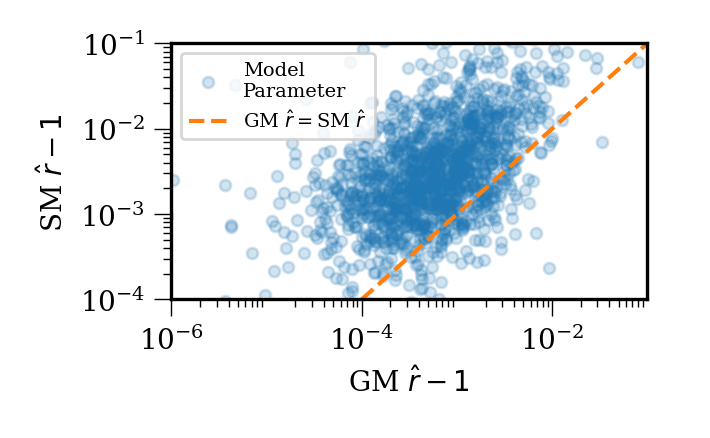}
    \caption{A scatter-plot showcasing the differences in the spread of rank-normalized-split-R-hat, $\hat{r}$, values obtained from the $\log_{10}\bm{\rho}$ and the white noise parameters analyzed by GM and SM. Each blue circle represents a single model parameter and the figure is obtained by combining the $\hat{r} - 1$ values of all model parameters for all pulsars and across all frequencies. As evident by the figure, GM is more capable at resulting in posteriors with a lower $\hat{r}$ level given the two hour time limit. The $\hat{r}$ values are estimated by dividing each Markov chain into two sub-chains and applying the rank-normalized-split-R-hat test \citep{arviz_2019} on it.}
    \label{GR Comp}
  \end{minipage}
\end{figure*}
\subsection{\label{sec:POS COMP}Comparison of posteriors}
For the sake of brevity, out of the thirty four pulsars of the NANOGrav 11 year data set, we have chosen to feature a GM vs SM posterior comparison plot for only PSR J$1713+0747$ as this pulsar has the longest observational baseline as well as the largest number of TOAs making it the most computationally expensive pulsar to analyze.
As shown in \autoref{POS Comp}, the two techniques yield consistent posteriors for both the red noise and the white noise model parameters for PSR J$1713+0747$ showcasing the robustness and the capability of GM to be implemented on real PTA data sets. The same consistency is also observed in all the remaining thirty three pulsars. For a quantification of the degree of consistency between the two sets of posteriors, refer to \autoref{HELL1} which highlights the differences in the output of GM and SM in the form of a histogram of \emph{Hellinger distance} \citep{Hell} \footnote{Hellinger distance is a measure of similarity between two probability distributions ranging from $0$ (identical distributions) to $1$ (disagreeing distributions). For two discrete probability distributions $p$ and $q$, the Hellinger distance $H$ is defined as $H=\frac{1}{\sqrt{2}} \sqrt{\sum\limits_{i}{{{\left( \sqrt{{{p}_{i}}}-\sqrt{{{q}_{i}}} \right)}^{2}}}}$, where $i$ ranges over the binned quantities of interest whose probability distribution is described by $p$ and $q$.} values across all pulsars. With the exception of a few white noise parameters, the Hellinger distances are concentrated between $0$ and $0.2$ indicating an adequate degree of consistency between the GM and the SM posteriors. We attribute the higher Hellinger distance values of some model parameters (especially the white noise parameters) to the differences in the level of convergence of the posteriors as GM is more successful at yielding converged posteriors than SM. Refer to \S\ref{sec:concomp} for a more detailed discussion.  
\subsection{\label{UnderSampleWN} The effect of using different number of MCMC steps in GM}
To obtain the white noise posteriors of \autoref{POS Comp}, 30 steps of a Metropolis Hasting algorithm for each step of Gibbs sampling has been implemented. The choice for the number of MCMC steps for each step of the Gibbs sampling depends on factors such as the number of TOAs, one's threshold and preferred measure of convergence for the posteriors as well as the efficiency of the type of MCMC algorithm used in the white noise parameter estimation. However, the red noise parameters' posteriors are not overly sensitive to this choice as the target distributions for $\rho_k$ parameters are all analytically determined prior to the start of sampling. To test the sensitivity of the red noise parameters to the choice for the number of MCMC steps for each step of Gibbs sampling, we have applied GM on all of the NANOGrav 11 year pulsars using only 5 steps of MCMC. As shown in \autoref{HELL2}, the estimated Hellinger distance values between the two sets of posteriors of $\log_{10}{\bm{\rho}}$ parameters are sufficiency low suggesting a weak degree of correlation between the red noise parameters' posteriors to the white noise parameters' if analyzed via GM. Nevertheless, our  current implementation of GM is adequately optimized to handle large number of MCMC steps without much of a sacrifice in the overall run-time of a single-pulsar analysis. 
\subsection{\label{sec:concomp}Comparison of convergence levels}
Despite resulting in consistent posteriors, SM and GM differ significantly in their state of convergence of the model parameters, especially those pertaining to the effective-sample-size (ESS). \autoref{ESS Comp} shows the spread of the ratio of ESS values (GM divided by SM) across all of the model parameters for every pulsar. As evident by \autoref{ESS Comp}, a significant majority of each pulsar's model parameters have higher ESS values when analyzed using GM as compared to SM. The average ESS ratio across all parameters and pulsars is 6. \autoref{ESS Comp} proves our claim about the high efficiency of GM. Additionally, the same observation can be made about the rank-normalized-split R-hat ($\hat{r}$) values calculated for both GM and SM posteriors for each pulsar. \autoref{GR Comp} points towards the higher state of convergence of a significant majority of the model parameters that were analyzed by GM.
\section{\label{sec:sims} Simulations}
Despite the successful implementation of GM on the NANOGrav 11 year data set, we have not tried to analyze the correlation content of the data set using the concepts discussed in \S\ref{sec:M3A} as the 11 year data set lacks a common correlated signal across pulsar pairs \citep{11yr}. For studying the correlations, we will dedicate future projects to the analysis of the NANOGrav 15 year \citep{d1} and the upcoming IPTA's DR3 data sets. Meanwhile, to explore the capability of the Fourier coefficients $\bm{a}$ in characterizing a common spatially-correlated signal, we make use of simulated PTA data sets.

\subsection{\label{sec:SIMDetail} Details of the simulations}
We have chosen two types of simulated data sets, referred to as SIM0 and SIM1, with 300 realizations for each type, to analyze in order to explore the capability of the Fourier coefficients $\bm{a}$ to characterize a common correlated signal. The two simulated data sets are identical in every aspect except the content of their spatially-uncorrelated intrinsic red noise: for SIM0, the log of the amplitude of the spatially-uncorrelated intrinsic red noise of each pulsar is randomly chosen from a uniform distribution between $10^{-16}$ and $10^{-14}$ while for SIM1 this range is between $10^{-14}$ and $10^{-13}$. For both data sets' pulsars, the spectral index of the spatially-uncorrelated intrinsic red noise follows a uniform distribution with lower and upper bounds of $0$ and $7$ respectively. Additionally, each data set has 90 pulsars uniformly scattered across the sky timed for 20 years with random timing cadences between 14 to 30 days. Furthermore, each data set contains $10$ microseconds of white Gaussian noise for each pulsar as well as a unique realization of a GWB with amplitude of $A_g = 2\times 10^{-15}$ and spectral index of $\gamma_g = 13/3$. Lastly, to employ GM on each data set, we keep the white noise parameters constant and use the same range of frequency-bins for all pulsars which is $\left\{ 1/{20\:\text{yrs}},\ 2/{20\:\text{yrs}},\ 3/{20\:\text{yrs}},\ 4/{20\:\text{yrs}},\ 5/{20\:\text{yrs}} \right\}$. 

It is worth mentioning that our intention is not about simulating realistic data sets and analyzing it with GM. We have already shown the capability of GM in single-pulsar analyses of real data sets. Our intention is to highlight what the Fourier coefficients can potentially reveal about an existing GWB signal, hence the reason behind our choices for the specific parameters of the two simulated data sets. Nonetheless, we have introduced very high levels of spatially-uncorrelated intrinsic red noise in the SIM1 data set (higher than what is observed in the real PTA data sets) as dealing with such processes is an extremely challenging part of GWB searches using PTAs whose impact on the correlation recovery using the Fourier coefficients is non-trivial. 
\begin{figure}
    \includegraphics[width=\linewidth]{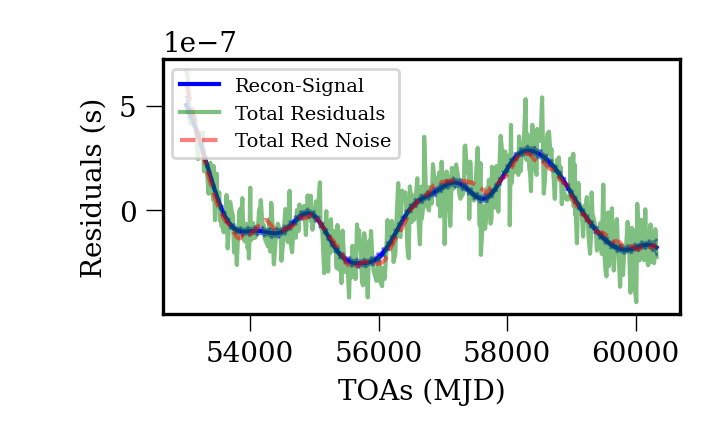}
    \caption{A comparison between postfit time series reconstruction using the Fourier coefficients obtained from GM (blue), the injected red noise time series (red), and the total residuals (green) for one of SIM0's pulsars. The reconstructed residuals are made by considering the entire posterior probability distribution of the recovered Fourier coefficients. As evident by the figure, the reconstructed post-fit red noise signal matches the underlying red noise signal closely.}
    \label{TimeRecon}
\end{figure}
\begin{figure*}
    \includegraphics[width=\linewidth]{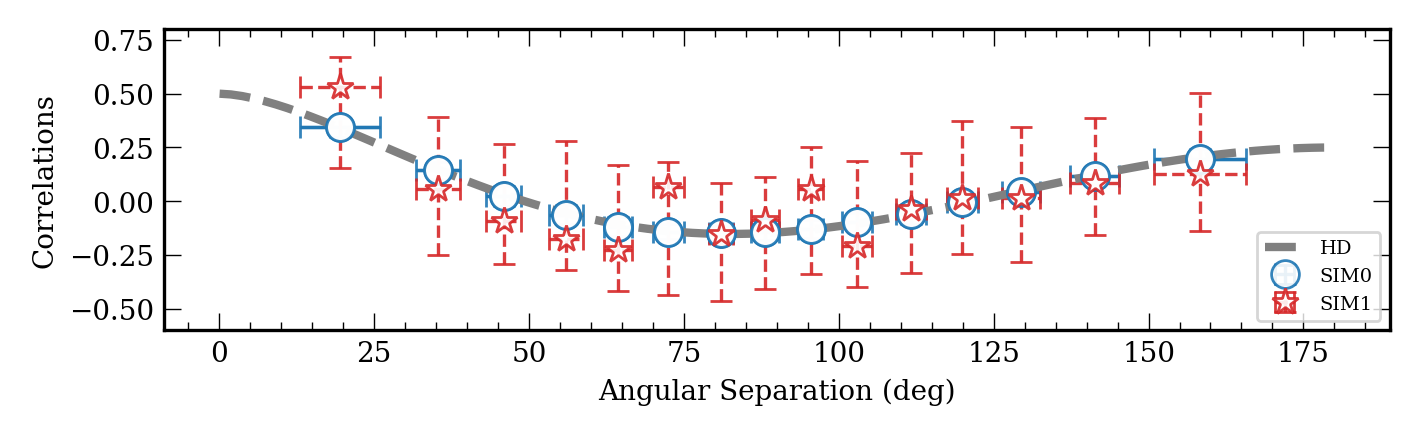}
    \caption{A plot depicting the reconstruction of the Hellings and Downs correlation (gray dashed curve) using GM's estimates of the Fourier coefficients obtained for both SIM0 (blue circles) and SIM1 (red stars) data sets. The reconstructions are the average over 300 realizations of both data sets. The error-bar of each point indicates the range between the 16th and the 84th percentiles over the 300 realizations. Remarkably, the recovery of the shape of the correlations is not affected significantly by the introduction of extremely high levels of spatially-uncorrelated red noise to each pulsar in SIM1.}
    \label{SimCorr}
\end{figure*}
\subsection{\label{sec:timerecon} Reconstruction of red noise signal using Fourier coefficients}
\begin{figure}
    \includegraphics[width=\linewidth]{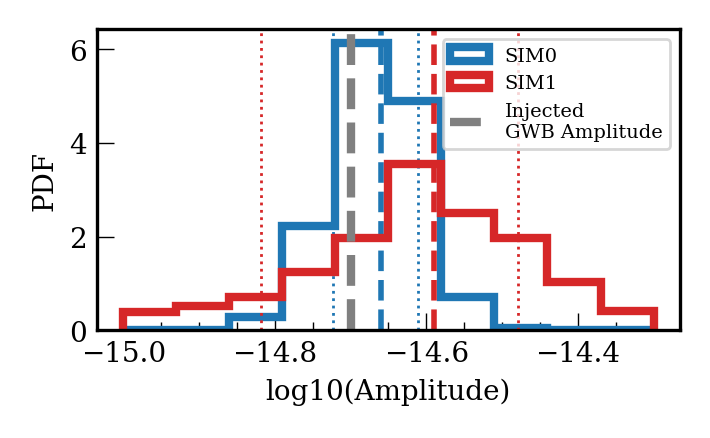}
    \caption{Two histograms comparing the distributions of the recovered common correlated signal between SIM0 (blue) and SIM1 (red) data set using the method provided in \S\ref{sec:M3A}. The blue and the red vertical lines indicate the 16th and the 84th percentiles (dotted lines) as well as the mean (dashed line) of each distribution. Each distribution is obtained by combining the estimates of the amplitude (\autoref{opt}) of the cross-correlated signal over 300 realizations. The injected GWB signal is indicated with a vertical dashed gray line. The figure suggests that the Fourier coefficients contain the right amount of information about the amplitude of the cross-correlated signal in the case of SIM0. In the case of SIM1, due to the existence of significantly higher non-GWB red noise power, the recovered GWB amplitude is more scattered.}
    \label{SimAmp}
\end{figure}
\begin{figure}
    \includegraphics[width=\linewidth]{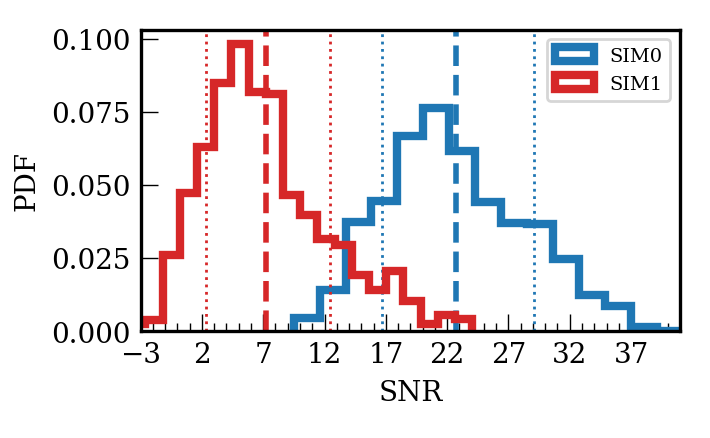}
    \caption{Two histograms comparing the distributions of the signal-to-noise ratio (SNR) between SIM0 (blue) and SIM1 (red) data set using the method provided in \S\ref{sec:M3A}. The blue and the red vertical lines indicate the 16th and the 84th percentiles (dotted lines) as well as the mean (dashed line) of each distribution. Each distribution is obtained by combining the estimates of the SNR (\autoref{SNR}) of the cross-correlated signal over 300 realizations. As expected, SIM1 data set exhibits a lower SNR due to containing a significantly higher non-GWB red noise power than the GWB power.}
    \label{SimSNR}
\end{figure}

The $\bm{a}$ coefficients are capable of reconstructing the red component of the timing residulas as suggested by \autoref{firsteq}. The reconstructed signal is \emph{pre-fit} and \emph{white-noise-free}. Once the reconstructed signal obtained by $Fa$ is fitted for the timing model parameters, it mirrors the underlying total \emph{post-fit} red noise signal in the data set closely. \autoref{TimeRecon} highlights this case for one of SIM0's pulsars. As suggested by the figure, the Fourier coefficients are capable of reconstructing the underlying red noise process of the total timing residuals. This fact allows the Fourier coefficients to be adequate replacement for the timing residuals in the frequency domain with the added benefit that one no longer needs to take into account a white noise process or be concerned with the complications of the timing model parameters when using the $\bm{a}$ coefficients in a subsequent analysis. In fact, the effects of the timing model parameters and the white noise levels are implicit in the posteriors for the Fourier coefficients obtained via GM. 

\subsection{\label{sec:crosscorr} Searching for correlations using Fourier coefficients}
To characterize the GWB signal in each of the realizations of SIM0 and SIM1, we use \autoref{lambda} and \autoref{lambdasigma} with $\hat{\mu}_I$ as the quantity representing the Fourier coefficients of each pulsar (see \autoref{bgivenrho}). Furthermore, since our goal is to  showcase the potential of the Fourier coefficients in revealing information about the GWB signal rather than outlining a complete and practical pipeline capable of fully characterizing a GWB signal, 
the weights $\varphi_I$ are set to the total red noise power that was used to generate the simulated data sets.

The shape of the correlation recovery is depicted in \autoref{SimCorr} for both simulated data sets. This shape is obtained by dividing the pulsar pairs of each realization into 15 different angular separation bins such that all bins have 267 pulsars pairs in them. Additionally, the average, 16th, and 84th percentiles (over the 300 realizations) of the correlations for each angular separation bin is computed and indicated in \autoref{SimCorr}. 
Furthermore, the histogram of the estimated amplitude $\hat{A_g}$ and signal-to-noise ratio of all the 300 realizations of each data set are stacked on top of each other (i.e., no averaging is performed) and presented in \autoref{SimAmp} and \autoref{SimSNR} respectively. The impact of introducing extreme levels of intrinsic spatially-uncorrelated red noise to the data set manifests itself in the form of lowering the signal-to-noise ratio and more scattered amplitude recovery. However, the shape of the correlations recovery remains remarkably close to the Hellings and Downs curve over many realizations.

\section{\label{sec:Discussion}Discussion and Future Work}
In this paper, we have shown that the Gibbs method (GM) is an efficient single pulsar Bayesian noise analysis technique capable of producing posteriors for the single-pulsar free-power-spectral-density and the white noise model parameters with convergence properties that are superior to those obtained using standard Bayesian methods (SM). GM is a robust and computationally efficient alternative to SM for future PTA noise analyses. Additionally, we have shown that the Fourier coefficients resulting directly from GM contain adequate information about the shape the cross-correlations signal through the use of simulations. In effect, GM produces the frequency domain representation of each pulsar's red noise signal, free of white noise and timing model parameters, hence providing all the necessary information to start performing subsequent GWB detection analyses exclusively in the frequency domain.


GM results in raw information in the frequency domain which may need to be processed further depending on the needs of the subsequent analyses. For instance, the astrophysical interpretation of a pulsar's red noise signal will require a more constrained model of the power spectral density than the free-spectrum model which could be achieved by fitting for the parameters of such a model using the output of GM (e.g., a power-law fit to the free-spectrum model) \cite{fitting}. Combined with the fitting utilities provided by \citet{fitting}, GM can become a powerful and efficient tool for use in the future PTA GWB detection analyses. 
\subsection{\label{sec:software}Software}
The GM code takes advantage of the functionalities provided by ENTERPRISE \citep{enterprise} and ENTERPRISE-extensions \citep{enterprise-ex}, and PTMCMC sampler \citep{PTMCMC}. The package Arviz \citep{arviz_2019} has been used for diagnosing MCMC chains. Python packages matplotlib \citep{plt} and plotly \citep{plotly} have been used for generating the figures in this paper.
\begin{acknowledgments}
We thank the anonymous referee for their helpful feedback which improved the quality of this work. We thank our colleagues in NANOGrav for fruitful discussions and feedback during the development of this technique. We thank Justin A. Ellis
 for his early work on this subject and the early version of the Gibbs method's code. The work of N.L., W.G.L, J.D.R, X.S., and S.R.T was supported by the NANOGrav NSF Physics Frontier Center awards \#2020265 and \#1430284. N.L. and X.S acknowledges the support from the George and Hannah Bolinger Memorial Fund, as well as the Larry W. Martin and Joyce B. O’Neill Endowed Fellowship in the College of Science at Oregon State University. S.R.T acknowledges support from NSF AST-2007993, and an NSF CAREER \#2146016. This work was conducted in part using the resources of the Advanced Computing Center for Research and Education (ACCRE) at Vanderbilt University, Nashville, TN. This work was performed in part at Aspen Center for Physics, which is supported by National Science Foundation grant PHY-2210452. J.D.R. acknowledges support from start-up funds from Texas Tech University.
\end{acknowledgments}

\appendix

\section{\label{sec:terminology}GWB Detection Terminology}
Most of the PTA noise analysis concepts have been developed over many years and scattered over many papers \citep{9Year,Ncal,OPTSTAT0,Gibbs0,ScalingLaws,TimeDomainPaper, book, Lentati:2012xb, Null, 10.1093/mnras/staa3411}. To help readers better understand the methods used in this paper, we define the necessary PTA noise analysis quantities and concepts in this section. Additionally, refer to \autoref{params} for a short description of the mathematical symbols used throughout this paper. 
\subsection{\label{sub-sec:Matrices}Basis matrices and their coefficients}
To model the contribution of any red noise process to the timing model residuals of a given pulsar, $\bm{r}_\text{Red}$, we employ a Fourier basis matrix and a vector of coefficients such that 
\begin{widetext}
\begin{align}
    \bm{r}_{\text{Red}} &= F\bm{a} \label{rred},\\
      F&=\left( \begin{matrix}
   \sin \left( 2\pi {{f}_{1}}{{t}_{1}} \right) & \cos \left( 2\pi {{f}_{1}}{{t}_{1}} \right) & \cdots  & \sin \left( 2\pi {{f}_{k}}{{t}_{1}} \right) & \cos \left( 2\pi {{f}_{k}}{{t}_{1}} \right)  \\
   \sin \left( 2\pi {{f}_{1}}{{t}_{2}} \right) & \cos \left( 2\pi {{f}_{1}}{{t}_{2}} \right) & \cdots  & \sin \left( 2\pi {{f}_{k}}{{t}_{2}} \right) & \cos \left( 2\pi {{f}_{k}}{{t}_{2}} \right)  \\
   \vdots  & \vdots  & \ddots  & \vdots  & \vdots   \\
   \sin \left( 2\pi {{f}_{1}}{{t}_{p}} \right) & \cos \left( 2\pi {{f}_{1}}{{t}_{p}} \right) & \cdots  & \sin \left( 2\pi {{f}_{k}}{{t}_{p}} \right) & \cos \left( 2\pi {{f}_{k}}{{t}_{p}} \right)  \\
\end{matrix} \right),\\
{\bm{a}^{T}}&=\left( a_{1}^{\sin },a_{1}^{\cos },\ldots ,a_{k}^{\sin },a_{k}^{\cos } \right),
\end{align}
\end{widetext}
for $t_p$ denoting the last measured TOA, $f_k$ denoting the kth considered frequency-bin, and $a^{\text{sin}}$ and $a^{\text{cos}}$ referring to the coefficients of $\sin$ and $\cos$ elements of the $F$ matrix respectively. 


To model the contribution of any linear timing model parameter to the timing residuals, $\bm{r}_{\text{T}}$, we use a basis matrix known as the \emph{timing-design-matrix} such that 
\begin{align}
      {\bm{r}_{T}}&=M \bm{\epsilon},  \\ 
  M&=\left( \begin{matrix}
   1 & {{t}_{1}} & t_{1}^{2} & \cdots   \\
   1 & {{t}_{2}} & t_{2}^{2} & \cdots   \\
   \vdots  & \vdots  & \vdots  & \cdots   \\
   1 & {{t}_{p}} & t_{p}^{2} & \cdots   \\
\end{matrix} \right).
\end{align}
While the first three columns of the design matrix models the quadratic spin down of all millisecond pulsars, the unspecified columns of the matrix are populated with various timing model contributions specific to each pulsar. 
Moreover, it is often convenient to project
the residuals onto a subspace orthogonal to the
timing model parameters, or in other words, to create \emph{fitted} timing residuals. The so-called $G$ matrix is a useful matrix obtained via singular-value-decomposition of the design-matrix constructed to perform the fitting:
\begin{align}
\begin{split}
        M&=US{{V}^{T}} \\ 
  {{G}_{xy}}&={{U}_{xy}}, 
\end{split}
\end{align}
where $x$ ranges from $1$ to $p$ (the number of TOAs) while $y$ ranges from $q$ to $p$ for $q$ being the total number of the linear timing model parameters.

To model the contribution of the white noise to the timing model residuals, $\bm{r}_{\text{w}}$, we consider a $m \times m$ identity matrix as the basis with the coefficients $\bm{n}$ such that
\begin{align}
      {\bm{r}_{w}}&= \bm{w}, \\ 
  {{w}_{i}}&\sim \text{Normal}\left( \text{mean} =0,\text{scale} =\sigma _{w_i} \right), \\ 
  \sigma _{w_i}&=\text{ef}_i\sqrt{\sigma _{i}^{2}+\text{e}{{\text{q}}_i^{2}}},
\end{align}
for $\sigma_i$ being the TOA error of observation $i$, and $\text{ef}$ and $\text{eq}$ being the usual EFAC and EQUAD parameters \citep{11yr}. Note that the Gausianity of the white noise is an assumption included in our all of our models.
\subsection{\label{sub-sec:PSD}Noise power-spectral-density modeling}
In this paper, we only consider one-sided power-spectral-densities (PSD). Most commonly for PTA noise analyses, the PSD is expressed  in two ways:
\begin{itemize}[leftmargin=*]
    \item[]\textbf{Power Law}: assuming the PSD to follow a simple power-law relation with amplitude $A$ and spectral index $\gamma$ as well as a reference frequency $f_\text{ref}$ across all frequency-bins
    \begin{align}
        {{P}}\left( f \right)&=\frac{A^{2}}{12{{\pi }^{2}}{{f}^{3}}}{{\left( \frac{f}{{{f}_{\text{ref}}}} \right)}^{3-\gamma }},\\
        \hat{P}\left( f \right)&=\frac{P\left( f \right)}{A^2}. \label{PL}
    \end{align}
The quantity $\hat{P}$ describes the shape of the spectrum and is used in \S\ref{sec:crosscorr}.
    \item[]\textbf{Free-spectrum}: allowing the PSD to have independent amplitude in each frequency-bin with normalization constant $T_\text{obs}$ equal to a fixed observation time. The observation time can either be the baseline of each pulsar or the baseline of the total PTA experiment. 
    \begin{align}
    {{P}}\left( {{f}_{k}} \right)={{T}_{\text{obs}}} \: \rho _{k}^{2} \label{rhodef}.
    \end{align}
\end{itemize} 

\subsection{\label{sub-sec:Cov}Covariance matrices}
The white noise covariance matrix $N$ plays a key role in posterior probability calculation of all model parameters. This matrix is modeled as
\begin{align} 
 N&=\text{diag}\left( {{\sigma}_{w_1}},...,{{\sigma}_{w_p}} \right).
 \end{align}
Note that the introduction of ECORR white noise parameter will complicate this picture. See chapter 7 of \citet{book} for more details.
Furthermore, the red process covariance matrix is obtained via the discretized form of the Wiener-Khinchin theorem
\begin{align}
   \left\langle {{r}_{\text{red}}}\left( {{t}_{i}} \right){{r}_{\text{red}}}\left( {{t}_{j}} \right) \right\rangle =& \left[F\varphi {{F}^{T}} \right ]_{ij}, \\
   \begin{split}
         \varphi =& \: \text{diag} \Big\{ {{P}_{\text{red}}}\left( {{f}_{1}} \right),{{P}_{\text{red}}}\left( {{f}_{1}} \right), \\& \ldots ,{{P}_{\text{red}}}\left( {{f}_{k}} \right), {{P}_{\text{red}}}\left( {{f}_{k}} \right) \Big\},
         \end{split}
\end{align}
where $P_\text{red}$ is the one-sided PSD of a red noise process and the diagonal matrix $\varphi$ is the matrix representation of that PSD. 
\section{\label{truncinvgamma} Truncated Inverse-gamma Distribution}
To obtain a truncated inverse-gamma distribution, we take advantage of \emph{inverse-transform sampling} method. However, first, we need to find a normalization factor, $\text{Norm}$, for the truncated inverse-gamma distribution defined between the lower bound $\rho_{\text{min}}$ and the upper bound $\rho_{\text{max}}$:
\begin{align}
 \beta_k &= \frac{\left( a_k \cdot a_k \right)}{2}, \\ 
\text{Norm}&={{\left[ \int_{{{\rho }_{\min }}}^{{{\rho }_{\max }}}{d{{\rho }_{k}}\left\{ \frac{\beta_k }{\rho _{k}^{2}}\exp \left( -\frac{\beta_k }{\rho_k } \right) \right\}} \right]}^{-1}} \nonumber \\ 
 &=\frac{\beta_k }{\exp \left( -\frac{\beta_k }{{{\rho }_{\max }}} \right)-\exp \left( -\frac{\beta_k }{{{\rho }_{\min }}} \right)} \label{norm}.
\end{align}
Note that the above process can be repeated for all frequency bins. \autoref{norm} allows for calculation of the cumulative distribution function (CDF), which in turn can be used to find a distribution for $\rho_k$ given a uniform random number $U$ defined between $0$ and $1$ based on inverse-transform sampling method. This yields the following as the target distribution for the $\rho_k$ parameters:
\begin{widetext}
\begin{align}
   p\left( \left. \rho_k  \right|a_k,r,n \right) =-\frac{\beta_k }{\ln \left\{ \text{exp}\left( -\frac{\beta_k }{{{\rho }_{\min }}} \right)U(0,1)\left[ \exp \left( -\frac{\beta_k }{{{\rho }_{\max }}} \right)-\exp \left( -\frac{\beta_k }{{{\rho }_{\min }}} \right) \right] \right\}}. \label{rhogivenbexplicit} 
\end{align}
\end{widetext}
\section{\label{FDOSNORM}Derivation of the Normalization Factor in \autoref{lambda}}
The choice of normalization in the denominator of \autoref{lambda} enforces the condition that the estimated cross correlations must yield GWB amplitude if averaged over many realizations as is shown below: 
\begin{align}
   {{\lambda }_{IJ}}&=\frac{{{p}_{\text{top}}}}{\text{Norm}}, \\ 
  {{p}_{\text{top}}}&=\sum\limits_{k}{\frac{{\bm{a}_{I}}}{{\varphi_{I}}}\frac{{\bm{a}_{J}}}{{\varphi_{J}}}{{{\hat{P}}}_{g}}}, \\ 
  \left\langle {{p}_{\text{top}}} \right\rangle &=\left\langle \sum\limits_{k}{\frac{{\bm{a}_{I}}}{{\varphi_{I}}}\frac{{\bm{a}_{J}}}{{\varphi_{J}}}{{{\hat{P}}}_{g}}} \right\rangle  \\ 
 &=\sum\limits_{k}{\left\langle \frac{{\bm{a}_{I}}}{{\varphi_{I}}}\frac{{\bm{a}_{J}}}{{\varphi_{J}}} \right\rangle {{{\hat{P}}}_{g}}}
 \end{align}
 \begin{align}
      &=\sum\limits_{k}{\frac{\left\langle {\bm{a}_{I}}{\bm{a}_{J}} \right\rangle {{{\hat{P}}}_{g}}}{{\varphi_{I}}{\varphi_{J}}}} \\ 
 &=\sum\limits_{k}{\frac{\left( {{\Gamma }_{IJ}}A_{g}^{2}{{{\hat{P}}}_{g}} \right){{{\hat{P}}}_{g}}}{{\varphi_{I}}{\varphi_{J}}}} \\ 
 &={{\Gamma }_{IJ}}A_{g}^{2}\sum\limits_{k}{\frac{\hat{P}_{g}^{2}}{{\varphi_{I}}{\varphi_{J}}}},
\end{align}
which makes $\text{Norm}=\sum\limits_{k}{\frac{\hat{P}_{g}^{2}}{{\varphi_{I}}{\varphi_{J}}}}$ consequently. 

Furthermore, when the quantity $\hat{\mu}$ of \autoref{mudef} is used in estimating the correlations following \autoref{lambda}, the normalization need to be re-estimated since $\left\langle {\bm{a}_{I}}\bm{a}_{J}^{T} \right\rangle \ne \left\langle {{\bm{\hat{\mu}} }_{I}}\bm{\hat{\mu}} _{J}^{T} \right\rangle$ for an average over many GWB realizations. The new normalization factor is found to be
\begin{align}
  \left\langle {{\bm{\hat{\mu}} }_{I}}\bm{\hat{\mu}} _{J}^{T} \right\rangle &=\left\langle {{\Sigma }_{I}}F_{I}^{T}D_{I}^{-1}{{F}_{I}}{\bm{a}_{I}}\bm{a}_{J}^{T}F_{J}^{T}{{\left( D_{J}^{-1} \right)}^{T}}{{F}_{J}}\Sigma _{J}^{T} \right\rangle \nonumber\\ 
 & ={{\Sigma }_{I}}F_{I}^{T}D_{I}^{-1}F\left\langle {\bm{a}_{I}}\bm{a}_{J}^{T} \right\rangle F_{J}^{T}{{\left( D_{J}^{-1} \right)}^{T}}{{F}_{J}}\Sigma _{J}^{T}.
\end{align}
\bibliography{main}
\nocite{*}
\end{document}